\documentclass[aip, jcp, amsmath, amssymb, reprint, longbibliography]{revtex4-1}
\usepackage{graphicx}% Include figure files
\usepackage{dcolumn}% Align table columns on decimal point
\usepackage{bm}% bold math
\usepackage{color}
\usepackage{gensymb}

\begin{document}

%\preprint{AIP/123-QED}

\title{Generalized entropy theory of glass-formation in fully flexible polymer melts}
%\thanks{Contribution of the National Institute of Standards and Technology - Work not subject to copyright in the United States.}

\author{Wen-Sheng Xu}
\email{wsxu0312@gmail.com}
\altaffiliation{Present address: Center for Nanophase Materials Sciences, Oak Ridge National Laboratory, Oak Ridge, TN 37831, USA}
\affiliation{James Franck Institute, The University of Chicago, Chicago, Illinois 60637, USA}

\author{Jack F. Douglas}
\email{jack.douglas@nist.gov}
\affiliation{Materials Science and Engineering Division, National Institute of Standards and Technology, Gaithersburg, Maryland 20899, USA}

\author{Karl F. Freed}
\email{freed@uchicago.edu}
\affiliation{James Franck Institute, The University of Chicago, Chicago, Illinois 60637, USA}
\affiliation{Department of Chemistry, The University of Chicago, Chicago, Illinois 60637, USA}

\date{\today}% It is always \today, today, but any date may be explicitly specified
\begin{abstract}
The generalized entropy theory (GET) offers many insights into how molecular parameters influence polymer glass-formation. Given the fact that chain rigidity often plays a critical role in understanding the glass-formation of polymer materials, the GET was originally developed based on models of semiflexible chains. Consequently, all previous calculations within the GET considered polymers with some degree of chain rigidity. Motivated by unexpected results from computer simulations of fully flexible polymer melts concerning the dependence of thermodynamic and dynamic properties on the cohesive interaction strength ($\epsilon$), the present paper employs the GET to explore the influence of $\epsilon$ on glass-formation in models of polymer melts with a vanishing bending rigidity, i.e., fully flexible polymer melts. In accord with simulations, the GET for fully flexible polymer melts predicts that basic dimensionless thermodynamic properties (such as the thermal expansion coefficient and isothermal compressibility) are universal functions of the temperature scaled by $\epsilon$ in the regime of low pressures. Similar scaling behavior is also found for the configurational entropy density in the GET for fully flexible polymer melts. Moreover, we find that the characteristic temperatures of glass-formation increase linearly with $\epsilon$ and that the fragility is independent of $\epsilon$ in fully flexible polymer melts, predictions that are again consistent with simulations of glass-forming polymer melts composed of fully flexible chains. Beyond an explanation of these general trends observed in simulations, the GET for fully flexible polymer melts predicts the presence of a positive residual configurational entropy at low temperatures, indicating a return to Arrhenius relaxation in the low temperature glassy state.
\end{abstract}

%\pacs{61.20.Ja, 82.70.Dd, 64.70.P-}
%\keywords{Suggested keywords}%Use showkeys class option if keyword display desired

\maketitle

\section{Introduction}

The generalized entropy theory (GET) provides an attractive theoretical framework for systematically studying changes in polymer glass-formation caused by alterations in important molecular parameters.~\cite{ACP_137_125} The predictive power of the GET is achieved by merging the lattice cluster theory (LCT)~\cite{ACP_103_335, JCP_141_044909} for the thermodynamics of polymer systems with the Adam-Gibbs (AG) theory,~\cite{JCP_43_139} which invokes a relationship between the structural relaxation time and the configurational entropy. Since real polymer materials possess a degree of chain rigidity, the GET was originally developed based on models of semiflexible polymers, an effect that has been successfully incorporated into the LCT in terms of a bending rigidity parameter $E_b$.~\cite{ACP_103_335} As a consequence, all previous calculations within the GET (e.g., see Refs.~\citenum{ACP_137_125, JCP_131_114905, JCP_138_234501, Mac_47_6990, Mac_48_2333, JCP_141_234903}) considered models of semiflexible polymers for $E_b/k_B\gg 0$ K with $k_B$ being Boltzmann's constant.

While the GET for semiflexible polymers has been useful for understanding experimental trends in glass-forming polymers,~\cite{ACP_137_125} most simulations are performed for models of polymers without explicit bending constraints (e.g., see Refs.~\citenum{NC_5_4163, JCP_140_204509, PNAS_112_2966, JCP_142_234907}). In particular, our recent simulations~\cite{Paper1, Paper2} for a coarse-grained bead-spring model of such ``flexible'' polymer melts~\cite{JCP_92_5057, PRA_33_3628} reveal trends in the dependence of the cohesive interaction strength ($\epsilon$) on certain properties of glass-formation that deviate from the earlier GET predictions for semiflexible polymers. Specifically, the simulations indicate that the characteristic temperatures of glass-formation increase nearly linearly with $\epsilon$,~\cite{Paper1, Paper2} while the GET for semiflexible polymers predicts a somewhat non-linear growth of the characteristic temperatures with $\epsilon$.~\cite{JCP_131_114905, JCP_138_234501, Mac_47_6990, Mac_48_2333, JCP_141_234903} More strikingly, the GET for semiflexible polymers indicates that increasing $\epsilon$ leads to a reduction in the fragility of glass-formation,~\cite{JCP_131_114905, JCP_138_234501, Mac_47_6990, Mac_48_2333, JCP_141_234903} while the fragility remains nearly unchanged with $\epsilon$ in simulations for fully flexible polymer melts.~\cite{Paper1, Paper2}

In order to better understand the unexpected simulation results for fully flexible polymer melts,~\cite{Paper1, Paper2} the present paper explores the influence of $\epsilon$ on glass-formation in the GET for models of polymer melts with a vanishing bending rigidity (i.e., $E_b/k_B=0$ K). It turns out that the GET for fully flexible polymer melts can successfully explain the trends observed in simulations. In accord with simulations,~\cite{Paper1} the GET for fully flexible polymer melts predicts that basic dimensionless thermodynamic properties (such as the thermal expansion coefficient and isothermal compressibility) are universal functions of the temperature scaled by $\epsilon$ in the regime of low pressures. More importantly, the GET for fully flexible polymer melts rationalizes the linear increase of the characteristic temperatures and the constancy of the fragility with $\epsilon$ by analyzing the influence of $\epsilon$ on the configurational entropy density, a central thermodynamic quantity in the GET that likewise displays a universal behavior when the temperature is scaled by $\epsilon$ for $E_b/k_B=0$ K in the regime of low pressures. Beyond an explanation of the general trends observed in simulations, the GET for fully flexible polymer melts further predicts the presence of a positive residual configurational entropy at low temperatures. From the viewpoint of the entropy theory,~\cite{JCP_43_139} this residual configurational entropy implies that the dynamics becomes Arrhenius in the glassy state for this particular set of polymer models with a vanishing bending rigidity.

The remainder of the present paper is organized as follows. Section II briefly introduces the GET and describes its extension to treat polymer glass-formation in the limit of a vanishing bending rigidity. Section III begins by deriving an analytical expression for the residual configurational entropy for fully flexible polymer melts, followed by a discussion of the influence of the cohesive interaction strength on basic thermodynamic properties, characteristic temperatures, and fragility of glass-formation. Section IV finally provides a summary of the present work.

\section{Generalized entropy theory of polymer glass-formation}

As described in the introduction, the basic idea of the GET is to combine the AG relation~\cite{JCP_43_139} with the LCT.~\cite{ACP_103_335, JCP_141_044909} In doing so, the GET provides a theoretical tool for investigating the influence of various molecular details on polymer glass-formation. This unique feature of the GET comes largely from the fact that the LCT provides an analytical expression for the specific Helmholtz free energy $f$ (i.e. $f=F/N_l$ with $F$ denoting the total Helmholtz free energy and $N_l$ being the total number of lattice sites) of a polymer melt,~\cite{ACP_103_335, JCP_141_044909}
\begin{equation}
	\beta f=\beta f^{mf}-\sum_{i=1}^6C_i\phi^i,
\end{equation}
where $\beta=1/(k_BT)$ with $T$ designating the absolute temperature. The term $\beta f^{mf}$ in Eq. (1) represents the zeroth-order mean-field contribution, which for a one-bending energy model reads
\begin{eqnarray}
	\beta f^{mf}=&&\frac{\phi}{M}\ln\left(\frac{2\phi}{z^{L}M}\right)+\phi\left(1-\frac{1}{M}\right)+ (1-\phi)\ln(1-\phi)\nonumber\\
	&&
	-\phi \frac{N_{2i}}{M}\ln(z_b),
\end{eqnarray}
where $\phi$ is the polymer filling fraction, $M$ is the molar mass defined by the total number of united atom groups in a single chain, $z$ represents the lattice coordination number, which is related to the spatial dimension via $z=2d$ for a $d$-dimensional hypercubic lattice, $L$ is the number of subchains, $z_b=(z_p-1)\exp(-\beta E_b)+1$ with $z_p=z/2$ and $E_b$ being the bending rigidity parameter, and $N_{2i}$ is the number of sequential bond pairs lying along the identical subchains.~\cite{ACP_103_335, JCP_141_044909} More specifically, the polymer filling fraction is defined as $\phi=N_pM/N_l$ with $N_p$ designating the number of polymer chains, $\epsilon$ describes the strength of attractive interactions between nearest neighbor united atom groups, and $E_b$ quantifies the strength of chain rigidity. Note that side chains with two or more united atom groups can have separate bending energies.~\cite{ACP_137_125} The coefficients $C_i$ $(i=1, ..., 6)$ in Eq. (1) are obtained by collecting terms corresponding to a given power of $\phi$,
\begin{equation}
	C_i=C_{i, 0}+C_{i, \epsilon}(\beta\epsilon)+C_{i, \epsilon^2}(\beta\epsilon)^2.
\end{equation}
While the coefficients $C_{i, 0}$, $C_{i, \epsilon}$, and $C_{i, \epsilon^2}$, tabulated in Ref.~\citenum{JCP_141_044909}, are generally functions of $z$, $T$, $E_b$, $M$, as well as a set of geometrical indices that reflect the size, shape, and bonding patterns of monomers, they become independent of $T$ for $E_b/k_B=0$ K.

Within the GET, glass-formation is the consequence of a broad (or ``rounded'') thermodynamic transition characterized by four characteristic temperatures.~\cite{ACP_137_125} These characteristic temperatures are evaluated from the LCT configurational entropy density $s_c$ (i.e., the configurational entropy per lattice site) at a constant pressure $P$. This $s_c$ exhibits a maximum $s_c^*$ as a function of $T$, thereby providing the high $T$ limit of $s_c$, an essential feature for use in the AG model.~\cite{JCP_43_139} Hence, the temperature corresponding to this maximum defines the onset temperature $T_A$, which signals the onset of non-Arrhenius behavior of the relaxation time. Following AG,~\cite{JCP_43_139} the GET identifies the ``ideal'' glass transition temperature $T_0$ as the temperature at which $s_c$ extrapolates to zero. The crossover temperature $T_c$, which separates two regimes of $T$ with qualitatively different dependences of the structural relaxation time on $T$, is also evaluated solely on the basis of $s_c$ as the temperature at which the second derivative of $Ts_c(T)$ with respect to $T$ vanishes. To compute the glass transition temperature $T_g$, the GET invokes the AG relation,~\cite{JCP_43_139}
\begin{equation}
	\tau_{\alpha}=\tau_\infty\exp[\beta\Delta\mu s_c^\ast/s_c(T)],
\end{equation}
where $\tau_\infty$ is the high temperature limit of the structural relaxation time, and $\Delta\mu$ is the activation free energy at high $T$. $\tau_\infty$ is set to be $10^{-13}$ s in the GET, which is a typical value for polymers.~\cite{PRE_67_031507} Motivated by the experimental data on the crossover temperature of various glass-formers,~\cite{PRE_67_031507} the GET estimates the high $T$ activation free energy from the empirical relation $\Delta\mu=6k_BT_c$, an approximation that neglects the entropic contribution to $\Delta\mu$.~\cite{ACP_137_125} The GET then identifies $T_g$ using the common empirical definition $\tau_{\alpha}(T_g)=100$ s, a condition that reflects the somewhat arbitrary cooling rate chosen by convention in experimental measurements of $T_g$. The fragility is also readily computed from the $T$ dependence of $\tau_{\alpha}$ in the GET. Illustrative computations for the characteristic temperatures and fragility of glass-formation appear in Refs.~\citenum{ACP_137_125} and~\citenum{ Mac_47_6990}.

Note that Freed~\cite{JCP_141_141102} has offered an extended transition state theory by accounting for the collective barrier crossing events observed in the simulations of glass-forming liquids, which provides a much firmer ground for the basic assumptions in the AG theory. Moreover, the GET has already been demonstrated to produce general results in agreement with a wide range of observations from experiments and simulations.~\cite{ACP_137_125} This extensive agreement thus demonstrates the general validity of the AG theory. Of course, serious limitations of the AG theory indicate that it does not provide the ultimate theory for glass-formation.

While the GET described above was originally developed for semiflexible polymer melts,~\cite{ACP_137_125} we find that fully flexible polymer melts display certain features that are not captured by the GET for semiflexible polymers. (By construction, the zeroth-order mean-field term of the free energy for models of fully flexible polymer melts~\cite{Mac_24_5076} is actually different from that for models of semiflexible polymer melts with $E_b/k_B=0$ K.~\cite{ACP_103_335} For simplicity, the semiflexible polymer melts with $E_b/k_B=0$ K are termed the fully flexible polymer melts in the present paper.) For instance, $s_c$ is found to no longer vanish for $E_b/k_B=0$ K at low $T$. However, we find that both $s_c^*$ and $T_c$ are still well identified even for $E_b/k_B=0$ K (see specific examples below), thereby providing necessary inputs (such as $s_c^*$ and $\Delta \mu$) for use in the AG theory. Hence, $T_g$ can still be computed by its common definition given above, and $T_0$ can now be obtained by fitting $\tau_{\alpha}$ in a range of $T$ above $T_g$ but below $T_c$ to the Vogel-Fulcher-Tammann (VFT) equation,~\cite{PZ_22_645, JACS_8_339, ZAAC_156_245}
\begin{equation}
	\tau_{\alpha}=\tau_0\exp\left(\frac{DT_0}{T-T_0}\right),
\end{equation}
where $\tau_0$ and $D$ denote the high $T$ limit of $\tau_{\alpha}$ and the fragility parameter quantifying the strength of the $T$ dependence of $\tau_{\alpha}$. For consistency, we also determine $T_0$ using the VFT fits even for $E_b/k_B>0$ K in the present paper.

\section{Results and discussion}

This section begins by deriving an analytical expression of the residual configurational entropy for fully flexible polymer melts, followed by a discussion of the influence of the cohesive interaction strength on basic thermodynamic properties. This section then discusses the influence of the cohesive interaction strength on the characteristic temperatures and fragility of glass-formation.

\subsection{Residual configurational entropy of fully flexible polymer melts from the lattice cluster theory}

The derivation of the residual configuratonal entropy $s_r$ follows the same procedure described in Ref.~\citenum{ACP_161_443} for semiflexible polymer melts, which is briefly presented here for fully flexible polymer melts (i.e., $E_b/k_B=0$ K). Since this quantity, $s_r$, usually occurs at very low $T$ where $\phi \rightarrow 1$, the free energy then simplifies in the limit of a vanishing bending rigidity parameter to,
\begin{eqnarray}
	\beta f=&&\frac{1}{M}\ln\left(\frac{2}{z^{L}M}\right)+\left(1-\frac{1}{M}\right)
	-\frac{N_{2i}}{M}\ln(z_p)\nonumber\\
	&&
	-\sum_{i=1}^6C_i.
\end{eqnarray}
However, the coefficients $C_{i, 0}$, $C_{i, \epsilon}$, and $C_{i, \epsilon^2}$ in $C_i$ are now independent of $T$, so the following relation holds,
\begin{equation}
	\beta \frac{\partial C_i}{\partial \beta}=C_{i, \epsilon}(\beta\epsilon)+2C_{i, \epsilon^2}(\beta\epsilon)^2.
\end{equation}
Consequently, the residual configurational entropy density $s_r=-\left. \partial f_r/\partial T\right|_{\phi}$ is derived as
\begin{eqnarray}
	s_r=&&-\frac{1}{M}\ln\left(\frac{2}{z^{L}M}\right)-\left(1-\frac{1}{M}\right)+\frac{N_{2i}}{M}\ln(z_p)\nonumber\\
	&&
	+\sum_{i=1}^6\left[C_{i,0} - C_{i, \epsilon^2}(\beta \epsilon)^2\right].
\end{eqnarray}
Taking advantage of the result that $\sum_{i=1}^6C_{i, \epsilon^2}=0$,~\cite{ACP_103_335, JCP_141_044909} $s_r$ appears for $E_b/k_B=0$ K in the following form,
\begin{eqnarray}
	s_r=&&-\frac{1}{M}\ln\left(\frac{2}{z^{L}M}\right)-\left(1-\frac{1}{M}\right)+\frac{N_{2i}}{M}\ln(z_p)\nonumber\\
	&&
	+\sum_{i=1}^6C_{i, 0}.
\end{eqnarray}
We see that $s_r$ depends only on monomer structure and chain length for fixed $z$. We examine below how these molecular factors influence $s_r$ by taking $z=6$.

\begin{figure}[tb]
	\centering
	\includegraphics[angle=0,width=0.45\textwidth]{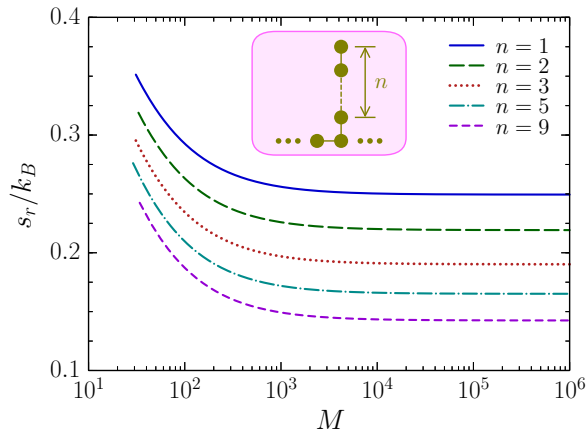}
	\caption{Residual configurational entropy density $s_r/k_B$ as a function of the molar mass $M$ for various side group lengths $n$ for fully flexible polymer melts with the structure of poly($n$-$\alpha$-olefins), as sketched in the figure.}
\end{figure}

\begin{figure}[tb]
	\centering
	\includegraphics[angle=0,width=0.45\textwidth]{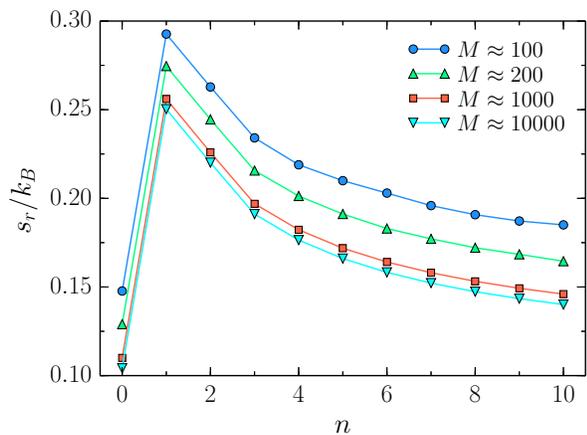}
	\caption{Residual configurational entropy density $s_r/k_B$ as a function of the side group length $n$ for various $M$ for fully flexible polymer melts. Due to the different numbers of united atom groups in a monomer for various $n$, the molar mass cannot exactly be identical to the indicated $M$ for all $n$.}
\end{figure}

Figure 1 displays the $M$ dependence of $s_r/k_B$ for various side group lengths $n$ for polymer melts with the structure of poly($n$-$\alpha$-olefins), as sketched in Fig. 1. In contrast to the results for semiflexible polymer melts with $E_b/k_B>0$ K,~\cite{ACP_161_443} $s_r$ is found to be always \textit{positive} for $E_b/k_B=0$ K, a result having significant implications for the dynamics of glass-forming polymers from the viewpoint of entropy theory,~\cite{ACP_137_125, JCP_43_139} as further discussed in Sec. III. Figure 1 also indicates that $s_r$ decreases with $M$ and saturates to a constant for sufficiently large $M$. Figure 2 examines the influence of side group length $n$ on $s_r$ for fixed $M$. We see that $s_r$ displays an abrupt increase when varying $n$ from $0$ to $1$, a result due to the sudden increase of branching points from linear chains to chains with the structure of poly(propylene) (PP). Increasing $n$ further from one to larger values leads to a drop in $s_r$, an effect similar to increasing $M$ for fixed $n$. There is substantial experimental evidence for a residual entropy in the glassy state (e.g., see Ref.~\citenum{JCP_132_124509} for discussion on this topic), and hence, the GET for fully flexible polymer melts provides theoretical support for these observations.

We next explore the influence of $\epsilon$ on polymer glass-formation for a fully flexible polymer melt with $E_b/k_B=0$ K using the GET described in Sec. II. For simplicity, all our calculations below are performed for polymer melts with the structure of PP and use the common parameters, $z=6$, $P=0.101~325$ MPa, $M=10~000$, and the cell volume parameter $V_{\text{cell}}=2.7^3$(\AA{})${^3}$, which is a parameter associated with the volume of a single lattice site in order to calculate $P$ in real units.~\cite{ACP_137_125} For comparison, results for a semiflexible polymer melt with $E_b/k_B=800$ K are also presented using the same parameter set.

\subsection{Basic thermodynamic properties}

Our discussion for glass-formation starts with the equation of state (EOS, i.e., the $T$ and $P$ dependence of the density) at a constant pressure, $P=-\left.\partial F/\partial V\right|_{N_p, T}$, where $V=N_l V_{\text{cell}}$ is the volume of the system. The explicit form of $P$ emerges in the LCT as, 
\begin{eqnarray}
	P = &&-\frac{1}{\beta V_{\text{cell}}}\left[\phi\left(1-\frac{1}{M}\right) + \ln(1-\phi) \right. \nonumber\\
	&&
	\left. + \sum_{i=1}^{6}(i-1)C_i\phi^i\right],
\end{eqnarray}
which must be solved numerically to produce $\phi$ as a function of $T$ for a given $P$. The EOS is a basic thermodynamic property that is readily obtained in experiments and simulations. Notice that $\phi$ is not a volume fraction but a filling fraction by definition in our lattice model. As described in previous work,~\cite{ACP_161_443} however, $\phi$ may be converted into a volume fraction if the polymer segments occupying the lattice sites are taken to be touching spheres located on the lattice sites. The polymer volume fraction is then readily shown to be $\pi \phi/6$ in our cubic lattice model, where the factor $\pi/6$ is simply the ratio of the volume of a sphere to the volume of a lattice site. We mention that Scher and Zallen have used the same argument to ``translate'' results for the filling fraction of the lattice model to continuum estimates of the percolation volume fraction in the context of modeling percolation on various types of lattices.~\cite{JCP_53_3759} For simplicity, our results for the EOS are presented only for $\phi (T, P)$.

We first focus on the case of $E_b/k_B=0$ K in Fig. 3. The lower inset to Fig. 3(a) displays $\phi$ as increasing greatly upon cooling for each $\epsilon$, and the upper inset to Fig. 3(a) highlights the non-linear growth of $\phi$ with $\epsilon$, in accord with simulations.~\cite{Paper1} The above features likewise hold for semiflexible melts with $E_b/k_B=800$ K [see the insets to Fig. 4(a)]. Figure 3(a) further shows that $\phi$ becomes a universal function of $T/\epsilon$, a scaling behavior that is also observed in simulations.~\cite{Paper1} It is readily seen from Eq. (10) that $\phi$ becomes exactly a universal function of $T/\epsilon$ for $E_b/k_B=0$ K in the limit of $P=0$ MPa because $C_i$ depends only on $T/\epsilon$ for a fixed molecular structure and chain length. However, this universal behavior is expected to disappear for large $P$ since the $T$ dependent term (i.e., $\beta PV_{\text{cell}}$) in Eq. (10) becomes more evident for larger $P$. We indeed found that the universal behavior shown in Fig. 3(a) for $P=0.101~325$ MPa clearly breaks down for $P \approx 10$ MPa (data not shown). Hence, this universal behavior holds only in the regime of low pressures, a result that is consistent with simulations,~\cite{Paper1} which are indeed performed under such conditions. The reason for the disappearance of the universal behavior at high $P$ is not known exactly at present. Increasing $P$ seems to enhance the coupling between the bond potential and the intermolecular potential, which probably underlies the non-universality of the thermodynamic properties at high $P$. We thus speculate that the inclusion of a torsional potential in the lattice model or a bending potential in the continuum model would lead to an even stronger coupling and non-universality.

\begin{figure}[tb]
	\centering
	\includegraphics[angle=0,width=0.45\textwidth]{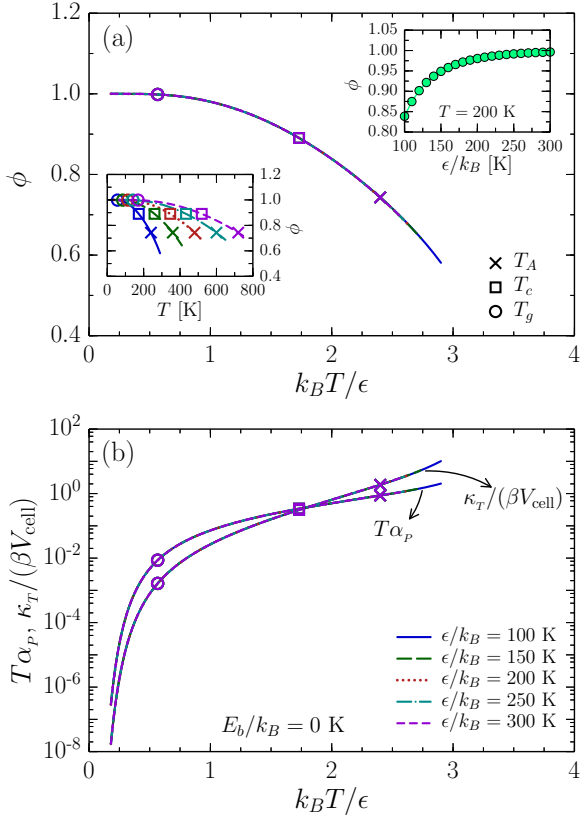}
	\caption{(a) Polymer filling fraction $\phi$ and (b) reduced thermal expansion coefficient $T\alpha_{_P}$ and reduced isothermal compressibility $\kappa_{_T}/(\beta V_{\text{cell}})$ as a function of $k_BT/\epsilon$ for various $\epsilon$. The lower and upper insets to (a) depict the $T$ dependence of $\phi$ for various $\epsilon$ and the $\epsilon$ dependence of $\phi$ at a fixed temperature of $T=200$ K, respectively. Pluses, squares, and circles indicate the positions of the onset temperature $T_A$, the crossover temperature $T_c$, and the glass transition temperature $T_g$, respectively. The calculations are performed for a melt of chains with the structure of poly(propylene) (PP) at a constant pressure of $P=0.101~325$ MPa, where the molar mass is $M=10~000$ and the bending rigidity parameter is $E_b/k_B=0$ K. Notice that the values of $\phi$, $T\alpha_{_P}$, and $\kappa_{_T}/(\beta V_{\text{cell}})$ at each characteristic temperature are independent of $\epsilon$.}
\end{figure}

\begin{figure}[tb]
	\centering
	\includegraphics[angle=0,width=0.45\textwidth]{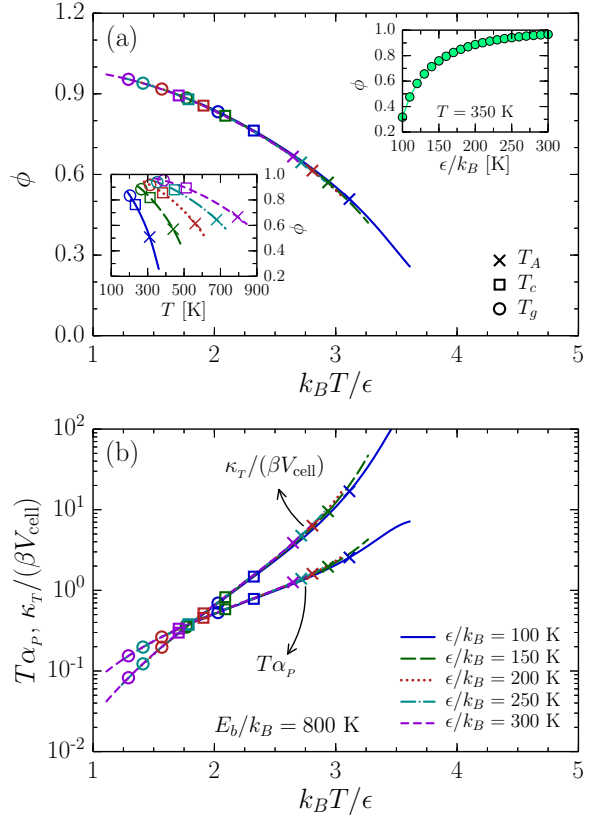}
	\caption{The same as in Fig. 3, except that the bending energy is $E_b/k_B=800$ K and that the temperature in the upper inset to (a) is fixed at $T=350$ K. Notice that the values of $\phi$, $T\alpha_{_P}$, and $\kappa_{_T}/(\beta V_{\text{cell}})$ at each characteristic temperature now depend on $\epsilon$.}
\end{figure}
	
It is also interesting to check whether ot not similar universal behaviors also appear for other basic thermodynamic properties. In particular, we consider the thermal expansion coefficient $\alpha_{_P}=(1/V)\left.(\partial V/\partial T)\right|_P$ and isothermal compressibility $\kappa_{_T}=-(1/V)\left.(\partial V/\partial P) \right|_T$, basic thermodynamic quantities that have been demonstrated to be of significance in polymer glass-formation~\cite{ACP_161_443} and that can readily be evaluated within the LCT,
\begin{eqnarray}
\alpha_{_P} = &&\frac{\sum_{i=1}^6\beta (\partial C_i/\partial \beta) (i-1)\phi^i + \beta PV_{\text{cell}}}{T\left[\phi^2/(1-\phi) + \phi/M - \sum_{i=1}^6 i(i-1)C_i \phi^i\right]},\nonumber\\
&&
\end{eqnarray}
and
\begin{eqnarray}
	\kappa_{_T} = \frac{\beta V_{\text{cell}}}{\phi^2/(1-\phi) + \phi/M - \sum_{i=1}^6i(i-1)C_i\phi^i}.
\end{eqnarray}
Motivated by our recent theoretical work based on the GET,~\cite{ACP_161_443} we consider the dimensionless thermal expansion coefficient and isothermal compressibility, which are given by $T\alpha_{_P}$ and $\kappa_{_T}/(\beta V_{\text{cell}})$, respectively, because these reduced quantities are found to be more useful than the dimensional quantities in the description of polymer glass-formation. Results for $T\alpha_{_P}$ and $\kappa_{_T}/(\beta V_{\text{cell}})$ are shown as a function of $T/\epsilon$ in Fig. 3(b) for $E_b/k_B=0$ K, where a master curve is apparent for each quantity. This feature is again consistent with simulations,~\cite{Paper1} where the reduced isothermal compressibility is obtained from the static structure factor. As noted in previous work,~\cite{Paper1} on approaching the glassy state, the liquid evidently shows an opposite tendency in comparison to approaching a liquid critical point, in the sense that the compressibility tends to vanish rather than diverge. Another noticeable feature shown in Fig. 3 is that the values of $\phi$, $T\alpha_{_P}$, and $\kappa_{_T}/(\beta V_{\text{cell}})$ at each characteristic temperature are independent of $\epsilon$.

For comparison, Fig. 4 displays the influence of $\epsilon$ on the same thermodynamic properties but for $E_b/k_B=800$ K. We see that the universal behaviors are still present, as evidenced by master curves between the dimensionless thermodynamic quantities and the scaled temperature $T/\epsilon$ in spite of small deviations at high $T$. However, one key difference emerges, i.e., the values of $\phi$, $T\alpha_{_P}$, and $\kappa_{_T}/(\beta V_{\text{cell}})$ at each characteristic temperature now depend on $\epsilon$.

\begin{figure}[tb]
	\centering
	\includegraphics[angle=0,width=0.45\textwidth]{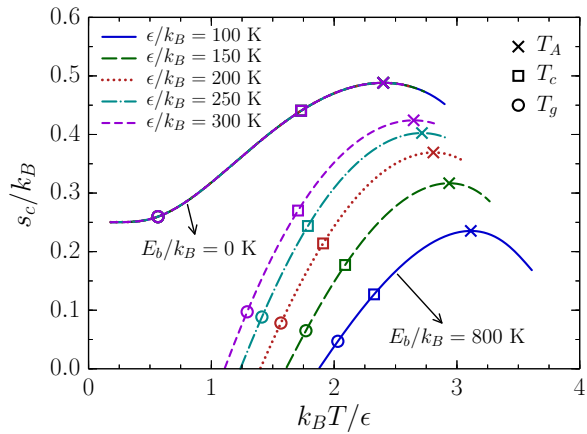}
	\caption{Configurational entropy density $s_c/k_B$ as a function $k_BT/\epsilon$ for various $\epsilon$ for $E_b/k_B=0$ K and $800$ K. Pluses, squares, and circles indicate the positions of $T_A$, $T_c$, and $T_g$, respectively. The calculations are performed for a melt of chains with the structure of PP at a constant pressure of $P=0.101~325$ MPa, where the chain length is $M=10~000$. Notice the absence of a universal curve describing the dependence of $s_c/k_B$ on $k_BT/\epsilon$ for $E_b/k_B=800$ K.}
\end{figure}

We now discuss the central quantity in the entropy theory of glass-formation, namely, the configurational entropy. In particular, the GET~\cite{ACP_137_125} demonstrates that the only meaningful configurational entropy for use in the AG theory~\cite{JCP_43_139} is the configurational entropy density $s_c$, i.e., the total configurational entropy per lattice site. This quantity, $s_c=-\left. \partial f/\partial T\right|_{\phi}$, is derived in the LCT for a semiflexible melt as,
\begin{eqnarray}
	\frac{s_c}{k_B} = &&-\beta f^{mf} - \beta \frac{\partial z_b}{\partial \beta} \frac{N_{2i}}{M} \frac{\phi}{z_b} + \sum_{i=1}^6\left(C_i-\beta \frac{\partial C_i}{\partial \beta}\right)\phi^i, \nonumber\\
	&&
\end{eqnarray}
which simplifies for $E_b/k_B=0$ K to the following equation,
\begin{eqnarray}
	\frac{s_c}{k_B} = &&-\frac{\phi}{M}\ln\left(\frac{2\phi}{z^{L}M}\right) - \phi\left(1-\frac{1}{M}\right) - (1-\phi)\ln(1-\phi) \nonumber\\
	&&
	+ \phi \frac{N_{2i}}{M}\ln(z_p) + \sum_{i=1}^6\left[C_{i,0} - C_{i, \epsilon^2}(\beta \epsilon)^2\right]\phi^i.
\end{eqnarray}
Equation (14) indicates that $s_c$ also becomes a universal function of $T/\epsilon$ if $\phi$ is a universal function of $T/\epsilon$.

Figure 5 displays $s_c/k_B$ as a function of $T/\epsilon$ for both a fully flexible melt and a semiflexible melt. The results of $s_c$ for $E_b/k_B=800$ K represent a typical picture for the $T$ dependence of $s_c$ within the GET for semiflexible polymers; i.e., $s_c$ displays a maximum at a high $T$ and vanishes at a low $T$. (The vanishing of $s_c$ might be an artifact of the high $T$ expansion in the LCT for semiflexible polymers at low $T$.~\cite{ACP_137_125, ACP_161_443}) In contrast to the thermodynamic properties discussed in Fig. 4, $s_c$ is not a universal function of $T/\epsilon$ for $E_b/k_B=800$ K. Moreover, the values of $s_c$ at each characteristic temperature clearly depend on $\epsilon$, a result that is relevant to understanding the dependence of $\epsilon$ on the fragility of glass-formation for semiflexible polymer melts. 

Turning to the case of $E_b/k_B=0$ K, a master curve evidently exists between $s_c/k_B$ and $T/\epsilon$, as expected, and the values of $s_c/k_B$ at each characteristic temperature are independent of $\epsilon$. This result, in conjunction with the AG relation given in Eq. (4), implies without further analysis that the strength of the $T$ dependence of $\tau_{\alpha}$ or the fragility is independent of $\epsilon$, while the characteristic temperatures of glass-formation increase in proportion to $\epsilon$. As discussed in Subsection III A, another interesting feature displayed by fully flexible polymer melts is that $s_c$ no longer vanishes but approaches a positive constant at low $T$. This prediction is consistent with the string model of glass-formation,~\cite{JCP_140_204509} which predicts a plateau for the average string length at low $T$, a quantity that is found to be inversely proportional to the configurational entropy.

The implications of a positive residual configurational entropy have recently been discussed in our previous work,~\cite{ACP_161_443} where the GET is employed to explore glass-formation in variable spatial dimension and where a positive residual configurational entropy is shown to exist for semiflexible polymer melts only above a critical dimension around $d=8$. The presence of a positive residual configurational entropy indicates that $\tau_{\alpha}$ does not diverge at any finite temperature and that the material is therefore a ``liquid'' at low $T$ from a mathematical standpoint. Hence, the Kauzmann entropy ``catastrophe'',~\cite{CR_43_219} implying that the configurational entropy would be negative, is avoided in the GET for fully flexible polymer melts. The residual configurational entropy at low $T$ further implies that structural relaxation becomes of an Arrhenius form, albeit with a higher activation energy than that which the fluid has at high $T$ above the onset temperature $T_A$. As a consequence, the relaxation times at low $T$ are much larger than those in the high $T$ regime, and by all practical measures, the polymer melt in the low $T$ regime can be considered to be a ``solid'' in a rheological sense. We are thus tempted to term the low $T$ state as being a type of a ``glass''. Recent experimental measurements~\cite{NC_4_1783, PRE_92_062304} suggest that relaxation in glass-forming materials generally approaches Arrhenius behavior at low $T$, as the GET predicts for fully flexible polymers.

\begin{figure}[tb]
	\centering
	\includegraphics[angle=0,width=0.45\textwidth]{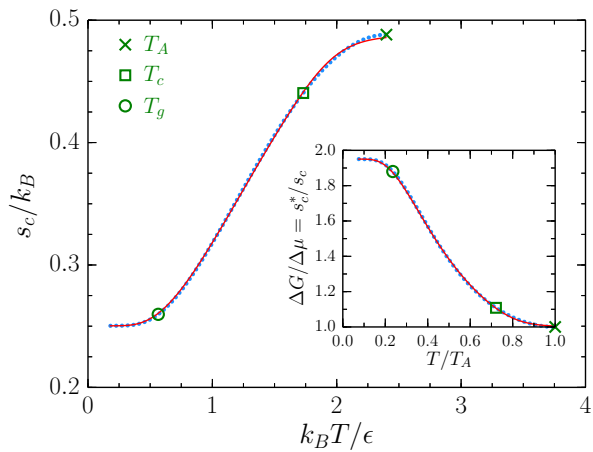}
	\caption{Configurational entropy density $s_c/k_B$ calculated from the GET (the dotted line) as a function $k_BT/\epsilon$ with $\epsilon/k_B=200$ K for $E_b/k_B=0$ K. The solid line is a fit to the empirical equation, $s_c = s_r + (s_c^* - s_r)[1 + \exp(a + b\epsilon\beta)]^c$, where the fitting parameter are determined to be $a=-13.88$, $b=28.24$, and $c=-8.72 \times 10^{-2}$. The residual configurational entropy and the high $T$ limit of $s_c$ are $s_r/k_B=0.2503$ and $s_c^*/k_B=0.4882$, respectively. The inset displays $\Delta G/\Delta \mu = s_c^*/s_c$ calculated from the GET (the dotted line) as a function of $T/T_A$ with the fit (the solid line) based on the above equation for $s_c$. Pluses, squares, and circles indicate the positions of the onset temperature $T_A$, the crossover temperature $T_c$, and the glass transition temperature $T_g$, respectively. These characteristic temperatures are determined to $T_A=479.9$ K, $T_c=346.1$ K, and $T_g=113.0$ K for this parameter set, respectively. The calculations are performed for a melt of chains with the structure of PP at a constant pressure of $P=0.101~325$ MPa, where the chain length is $M=10~000$.}
\end{figure}

We further discuss the sigmoidal variation of $s_c$ with $T$ for $E_b/k_B=0$ K. We find that the $T$ dependence of $s_c$ can be reasonably described by the following empirical equation,
\begin{eqnarray}
	s_c = s_r + (s_c^* - s_r)[1 + \exp(a + b\epsilon\beta)]^c,
\end{eqnarray}
where the constants $a$, $b$, and $c$, summarized in the caption of Fig. 6, are obtained by numerically fitting $(s_c-s_r)/(s_c^* - s_r)$ as a function of $T$ since this quantity varies from zero to unity as $T$ changes, thereby motivating the fitting functional form, $[1 + \exp(a + b\epsilon\beta)]^c$. Specifically, this fitting function is inspired by recent work based on the string model of glass-formation,~\cite{JCP_140_204509} which establishes a relation between emergent elasticity and cooperative motion in polymeric glass-forming liquids~\cite{PNAS_112_2966} and a quantitative inverse relation between the extent of cooperative motion and the configurational entropy,~\cite{JCP_138_12A541} as suggested by AG.~\cite{JCP_43_139} The relation between the activation free energy and material stiffness is further found to qualitatively accord with the shoving model of Dyre and co-workers.~\cite{PRB_53_2171} Based on these observations and arguments, we obtain Eq. (15) by first invoking a simple two-state model of the shear modulus $G$ of amorphous materials,~\cite{SM_6_3548} $G / G(T=0) = 1/ [1 + \exp(a + b \beta)]$. Equation (15) is further motivated by the observation that the activation free energy for relaxation scales approximately as a positive power of $G$ in a model glass-forming material.~\cite{JSM_5_054048} While these arguments are admittedly intuitive, Eq. (15) turns out to provide a good approximation for the $T$ dependence of $s_c$ calculated from the GET for fully flexible polymer melts. 

The GET then provides the following form for $\tau_{\alpha}$ for fully flexible polymer melts,
\begin{eqnarray}
\tau_{\alpha} = && \tau_\infty\exp\left\{\frac{\beta \Delta \mu}{\delta s + (1-\delta s)[1 + \exp(a + b\epsilon\beta) ]^c}\right\}, \nonumber\\
&&
\end{eqnarray}
where $\delta s=s_r/s_c^*$ is the ratio of the activation energy in the high $T$ regime to that in the low $T$ regime, similar to a quantity in the string model~\cite{JCP_140_204509} and the Doremus model~\cite{JAP_92_7619} of glass-formation. This approximation works in the entire range of $T$ below $T_A$ and thus may be useful in organizing relaxation data for polymers. 

The inset to Fig. 6 shows $\Delta G(T)/\Delta \mu = s_c^*/s_c(T)$ calculated from the GET as a function of $T/T_A$ for $\epsilon/k_B=200$ K, along with the fit based on Eq. (15). Here, $\Delta G(T)$ is the activation free energy, and the quantity, $s_c^*/s_c(T)$, thus describes the ratio of the activation free energy at $T$ to that at $T_A$ (or in the high $T$ limit). Evidently, the activation free energy increases sigmoidally with decreasing $T$, so the GET for fully flexible polymer melts does not predict any diverging relaxation time at any finite $T$, consistent with the string model of glass-formation.~\cite{JCP_140_204509} 

\subsection{Characteristic temperatures and fragility of glass-formation}

We now focus on the dependence of the characteristic temperatures on $\epsilon$ both for the fully flexible and semiflexible melts. As implied from the dependence of $s_c$ on $T/\epsilon$ discussed in Subsection III B, the characteristic temperatures of glass-formation should increase with $\epsilon$ in a linear fashion for the fully flexible melt. This linear behavior is clearly seen in Fig. 7(a), and thus, the GET for fully flexible melts indeed predicts the linear growth of the characteristic temperatures, as observed in simulations.~\cite{Paper1, Paper2} More specifically, the variation of each characteristic temperature $T_x$ ($x=A, c, g, \text{or}~ 0$) with $\epsilon$ follows the simple relation, $T_x=H_x\epsilon$, where the fitting parameter $H_x$ is given in the caption of Fig. 6. Meanwhile, Fig. 7(b) indicates that the growth of $T_x$ with $\epsilon$ becomes somewhat non-linear as $E_b$ increases, a result that is also revealed in previous calculations within the GET.~\cite{JCP_131_114905, JCP_138_234501, Mac_47_6990, Mac_48_2333, JCP_141_234903} While this non-linear behavior is more pronounced for $T_g$ and $T_0$ than for $T_A$ and $T_c$, we find that the dependence of all characteristic temperatures on $\epsilon$ for $E_b/k_B=800$ K can be described by the simple empirical relation, $T_x=(u_x + v_x\epsilon)/(1 + w_x\epsilon)$, where the fitting parameters $u_x$, $v_x$, and $w_x$ are provided in the caption of Fig. 7.

\begin{figure}[tb]
	\centering
	\includegraphics[angle=0,width=0.475\textwidth]{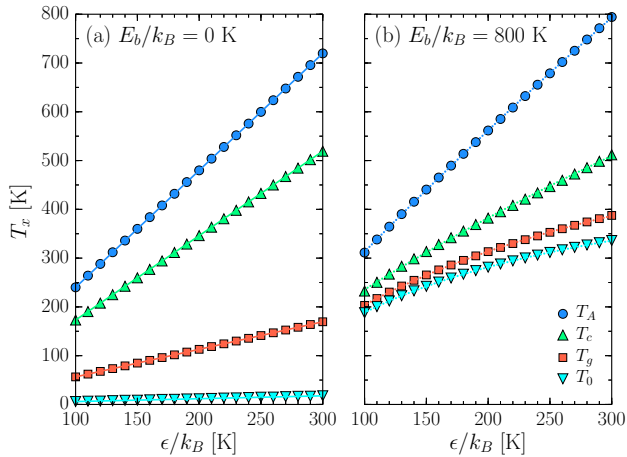}
	\caption{Onset temperature $T_A$, crossover temperature $T_c$, glass transition temperature $T_g$, and ideal glass transition temperature $T_0$ as a function of $\epsilon$ for (a) $E_b/k_B=0$ K and (b) $E_b/k_B=800$ K. Solid lines in (a) indicate the fits of data to the equation $T_x=H_x\epsilon/k_B$ ($x=A, c, g, \text{or}~ 0$), where the fitting parameters are determined to be $H_A=2.40$, $H_c=1.73$, $H_g=0.57$, and $H_0=5.90 \times 10^{-2}$. Dotted lines in (b) indicate the fits of data to the empirical equation $T_x=(u_x + v_x\epsilon/k_B)/(1 + w_x\epsilon/k_B)$ ($x=A, c, g, \text{or}~ 0$), where the fitting parameters are determined to be $(u_A, v_A, w_A)=(42.67, 2.81, 3.91\times 10^{-4})$, $(u_c, v_c, w_c)=(63.49, 1.92, 8.56\times 10^{-4})$, $(u_g, v_g, w_g)=(21.46, 2.51, 3.33\times 10^{-3})$, and $(u_0, v_0, w_0)=(-7.92, 3.00, 5.53\times 10^{-3})$. The calculations are performed for a melt of chains with the structure of PP at a constant pressure of $P=0.101~325$ MPa, where the molar mass is $M=10~000$.}
\end{figure}

\begin{figure}[tb]
	\centering
	\includegraphics[angle=0,width=0.45\textwidth]{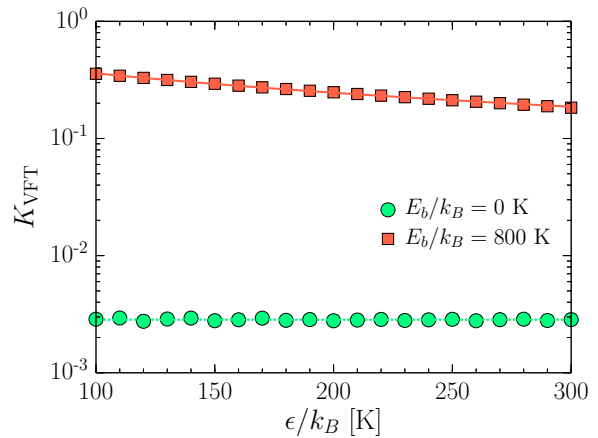}
	\caption{VFT fragility parameter $K_{\text{VFT}}$ as a function of $\epsilon$ for $E_b/k_B=0$ K and $800$ K, respectively. The dotted line indicates the average of $K_{\text{VFT}}$ over $\epsilon$ for $E_b/k_B=0$ K. The solid line represents a fit of data for $E_b/k_B=800$ K to the empirical equation, $K_{\text{VFT}}=k_0/(1+k_1\epsilon/k_B)$, where the fitting parameters are determined to be $k_0=0.67$ and $k_1=8.64 \times 10^{-3}$. The calculations are performed for a melt of chains with the structure of PP at a constant pressure of $P=0.101~325$ MPa, where the molar mass is $M=10~000$.}
\end{figure}

Figure 8 displays the $\epsilon$ dependence of the VFT fragility for both the fully flexible and semiflexible melts, where we consider $K_{\text{VFT}} \equiv 1/D$, a quantity that conveniently grows as the fragility increases. $K_{\text{VFT}}$ for fixed $\epsilon$ is evidently much smaller for $E_b/k_B=0$ K than for $E_b/k_B=800$ K, a result that has been explained by the GET in terms of packing frustration.~\cite{ACP_137_125} More rigid chains exhibit larger packing frustration, and hence, the melt composed of more rigid chains becomes more fragile. The same mechanism likewise accounts for the change of $K_{\text{VFT}}$ with $\epsilon$ for $E_b/k_B=800$ K, for which a trend of decreasing $K_{\text{VFT}}$ with increasing $\epsilon$ is observed in Fig. 8. The variation of $K_{\text{VFT}}$ with $\epsilon$ is described by the simple empirical relation, $K_{\text{VFT}}=k_0/(1+k_1\epsilon/k_B)$, where the fitting parameters $k_0$ and $k_1$ are given in the caption of Fig. 8. However, this mechanism associated with packing frustration appears unable to explain the variation of $K_{\text{VFT}}$ with $\epsilon$ for $E_b/k_B=0$ K, for which packing should become more efficient for larger $\epsilon$ but the fragility remains unchanged. However, this result is in agreement with simulations for fully flexible polymer melts,~\cite{Paper1, Paper2} and the constancy of $K_{\text{VFT}}$ with $\epsilon$ for $E_b/k_B=0$ K can be expected from the universal variation of $s_c$ with $T/\epsilon$, as discussed in Subsection III B.

\section{Summary}

Despite the long-recognized fact that monomer molecular and chemical structures directly determine the molecular parameters (such as cohesive energy and chain stiffness) governing the physical properties of polymers, a complete microscopic understanding of how important molecular parameters influence polymer glass-formation remains elusive. The GET~\cite{ACP_137_125} provides a convenient vehicle for systematically studying changes in polymer glass formation induced by alterations of key molecular parameters such as cohesive energy. However, our recent simulations~\cite{Paper1, Paper2} for a coarse-grained bead-spring model of flexible polymer melts reveal certain results for the dependence of the cohesive interaction strength ($\epsilon$) on glass-formation that are not expected from the GET for semiflexible polymers. For instance, simulations indicate that the characteristic temperatures of glass-formation increases with $\epsilon$ in a nearly linear fashion,~\cite{Paper1, Paper2} while the GET for semiflexible polymers predicts a somewhat non-linear growth of the characteristic temperatures with $\epsilon$.~\cite{JCP_131_114905, JCP_138_234501, Mac_47_6990, Mac_48_2333, JCP_141_234903} More strikingly, the GET for semiflexible polymers indicates that an increase in $\epsilon$ leads to a reduction in the fragility of glass-formation,~\cite{JCP_131_114905, JCP_138_234501, Mac_47_6990, Mac_48_2333, JCP_141_234903} but the fragility remains nearly unchanged with $\epsilon$ in simulations of glass-forming polymer melts composed of fully flexible chains.~\cite{Paper1, Paper2}

In order to better understand the simulation results,~\cite{Paper1, Paper2} the present paper employs the GET to explore the influence of $\epsilon$ on glass-formation in models of polymer melts with a vanishing bending rigidity. It turns out that this extension can successfully explain the trends observed in simulations. In accord with simulations,~\cite{Paper1} the GET for fully flexible chains predicts that basic dimensionless thermodynamic properties (such as polymer filling fraction, thermal expansion coefficient, and isothermal compressibility) are universal functions of the temperature $T$ scaled by $\epsilon$. These universal behaviors are predicted to occur only in the regime of low pressures, however. More importantly, the GET for fully flexible polymer melts rationalizes the linear increase of the characteristic temperatures and the constancy of the fragility with $\epsilon$ by analyzing the influence of $\epsilon$ on the configurational entropy density, a central thermodynamic quantity in the entropy theory that likewise displays a universal behavior when the temperature is scaled by $\epsilon$. Beyond an explanation of the general trends observed in simulations, the GET for fully flexible polymer melts further predicts the presence of a positive residual configurational entropy at low temperatures. From the viewpoint of entropy theory,~\cite{JCP_43_139} this residual configurational entropy implies that the dynamics becomes Arrhenius at low temperatures for the polymer models with a vanishing bending rigidity. This prediction is consistent with experimental measurements.~\cite{NC_4_1783, PRE_92_062304} We also note that the GET for fully flexible polymer melts predicts that the structural relaxation time likewise becomes a unique function of $T/\epsilon$, a behavior that somehow differs from simulations.~\cite{Paper2} This difference between the theory and simulations is not well understood at present, however.

In summary, we demonstrate that neglecting chain stiffness in the GET leads to a model that captures some of the observed aspects of real glass-forming materials, such as a residual configurational entropy, return to Arrhenius relaxation at low temperatures, etc. Introducing chain stiffness into the GET leads to a coupling of chain stiffness with the cohesive interactions, resulting in many interesting effects, such as the dependence of the fragility on $\epsilon$ and the non-linear variation of the characteristic temperatures with $\epsilon$, but the high temperature expansion for semiflexible polymer melts clearly leads to large errors at low temperatures. For example, the vanishing of the configurational entropy found in lattice models of polymers going back to the first calculations of polymer glass-formation by Gibbs and DiMarzio~\cite{JCP_28_373} is probably an artifact of this approximation. Future work is needed to simulate semiflexible chains to test the GET and to further understand the physical coupling between chain stiffness and cohesive interactions in glass-forming polymer melts.

Finally, let us make some remarks on the entropy ``catastrophe''. In particular, Gibbs and DiMarzio~\cite{JCP_28_373} predicted an entropy catastrophe based on a high temperature expansion of the torsional contribution to the configurational entropy of the polymer fluid, and they identified this thermodynamic condition with an ``ideal glass transition''. DiMarzio never accepted the arguments of AG,~\cite{JCP_43_139} and hence, the lattice model of polymers was not used to make quantitative estimates of structural relaxation times. Gujrati and Goldstein~\cite{JCP_74_2596} are the first to demonstrate that the Gibbs-DiMarzio model violated rigorous bounds on the entropy of polymer melts, clearly bringing the validity of this model into question, despite its tremendous empirical success in identifying essential trends between the experimental $T_g$ and the ideal glass transition temperature calculated from the lattice model. Issues with the vanishing of the configurational entropy predicted by both mean-field polymer and spin models have also been discussed in Ref.~\citenum{ACP_137_125}. The simulation work of Wolfgart \textit{et al.}~\cite{PRE_54_1535} clarified the situation somewhat by suggesting that the configurational entropy approaches a constant positive value at low temperatures in a bond fluctuation lattice model of flexible polymers, but the validity of these results has remained a question because of the difficulty of estimating thermodynamic properties reliably at such low temperatures. The idea of a finite configurational entropy in the glassy state remains controversial, a reason that makes our present results so interesting.

\begin{acknowledgments}
This work is supported, in part, by the National Science Foundation (NSF) Grant No. CHE-1363012.
\end{acknowledgments}

%\appendix

%\bibliographystyle{apsrev4-1}
\bibliography{refs}% Produces the bibliography via BibTeX.

\end{document}